\documentclass[10pt, preprint,  amsmath, amssymb]{article}
\usepackage{color}

\usepackage[affil-it]{authblk}

\usepackage{graphicx}
\usepackage{dcolumn}
\usepackage{bm}
\usepackage{siunitx}
\usepackage{amsmath}
\usepackage{amsfonts}
\usepackage{amssymb}

\newcommand{\ve}[1]{\ensuremath{{\mathbf{#1}}}}

\begin{document}

\title{A highly stable atomic vector magnetometer based on free spin precession}

\author{
 {S.~Afach,$^{1,2,3,12}$}
 {G.~Ban,$^4$}
 {G.~Bison,$^{1,13}$}
 {K.~Bodek,$^5$}
 {Z.~Chowdhuri,$^{1}$}
 {Z.~D.~Gruji\'c,$^{6}$}
 {L.~Hayen,$^{7}$}
 {V.~H\'{e}laine,$^{4}$}
 {M.~Kasprzak,$^{6}$}
 {K.~Kirch,$^{1,2  }$}
 {P.~Knowles,$^{6,8}$}
 {H.-C.~Koch,$^{6,9}$}
 {S.~Komposch,$^{1,2}$}
 {A.~Kozela,$^{10}$}
 {J.~Krempel,$^{2}$}
 {B.~Lauss,$^{1}$}
 {T.~Lefort,$^{4}$}
 {Y.~Lemi\`ere,$^{4}$}
 {A.~Mtchedlishvili,$^{1}$}
 {O.~Naviliat-Cuncic,$^{4}$}
 {F.~M.~Piegsa,$^{2}$}
 {P.~N.~Prashanth,$^{1}$}
 {G.~Qu\'em\'ener,$^{4}$}
 {M.~Rawlik,$^{5,2}$}
 {D.~Ries,$^{1,2}$}
 {S.~Roccia,$^{11}$}
 {D.~Rozpedzik,$^{5}$}
 {P.~Schmidt-Wellenburg,$^{1}$}
 {N.~Severjins,$^{7}$}
 {A.~Weis,$^{6}$}
 {E.~Wursten,$^{7}$}
 {G.~Wyszynski,$^{5}$}
 {J.~Zejma,$^{5}$} and
 {G.~Zsigmond$^{1}$}
 }

\affil{\small
 {$^1$Paul Scherrer Institute, 5232 Villigen PSI, Switzerland}\\
 {$^2$ETH Z\"urich, Institute for Particle Physics, 8093 Z\"urich, Switzerland}\\
 {$^3$Hans Berger Department of Neurology, Jena University Hospital, Jena, Germany}\\
 {$^4$Laboratoire de Physique Corpusculaire, Caen, France}\\
 {$^5$M.~Smoluchowski Institute of Physics, Jagiellonian University, Cracow, Poland}\\
 {$^{6}$University of Fribourg, Switzerland}\\
 {$^{7}$Katholieke Universiteit, Leuven, Belgium}\\
 {$^{8}$Present address: LogrusData, Rilkeplatz 8, Vienna, Austria}\\
 {$^{9}$Institut f\"ur Physik, Johannes-Gutenberg-Universit\"at, Mainz, Germany}\\
 {$^{10}$Henryk Niedwodnicza\'nski Institute for Nuclear Physics, Cracow, Poland}\\
 {$^{11}$Centre de Sciences Nucl\`eaires et de Sciences de la Mati\`ere, Orsay, France}\\
 {$^{12}$samer.afach@psi.ch}\\
 {$^{13}$georg.bison@psi.ch}
}

\maketitle

\begin{abstract}
We present a magnetometer based on optically pumped Cs atoms that
measures the magnitude and direction of a \SI{1}{\micro T} magnetic
field.
Multiple circularly polarized laser beams were used to probe the free
spin precession of the Cs atoms.
The design was optimized for long-time stability and achieves a scalar
resolution better than \SI{300}{fT} for integration times ranging from
\SI{80}{ms} to \SI{1000}{s}.
The best scalar resolution of less than \SI{80}{fT} was reached with
integration times of 1.6 to \SI{6}{s}.
We were able to measure the magnetic field direction with a resolution
better than \SI{10}{\micro rad} for integration times from \SI{10}{s}
up to \SI{2000}{s}.
\end{abstract}

\section{Introduction}

Magnetometers using optical pumping (OPM) of atomic media were
pioneered in the early 1960s \cite{Bloom1962}.
Since then, many OPM varieties \cite{Budker2007} have been developed
for diverse applications, e.g., mapping the geo-magnetic field or
detection of the bio-magnetic field emanating from the human heart
\cite{Bison2003, Wyllie20122247} and brain
\cite{Xia2006AMMEG1,Sander2012}.
In fundamental science, OPMs monitor the magnetic field in precision
magnetic resonance experiments searching for electric dipole moments
(EDM) \cite{Altarev1996, Knowles2009nEDMMagnetometer} and Lorentz
invariance tests \cite{Smiciklas2011, Peck2012PRA}.
The neutron EDM (nEDM) search sets stringent constraints on theories
proposing extensions beyond the standard model of particle physics.
Experimental sensitivity to a nEDM depends directly on the control
and measurement of the magnetic field in the experiment,
a task of particular challenge since (currently) the field must be
known over volumes on the order of 20~l for times of hundreds of seconds.
Herein, we present an OPM combining long-term stability with high
statistical sensitivity, and including vector information.
The OPM is designed to serve in an array of such sensors to
form an auxiliary magnetometer system monitoring the stability and
uniformity of the magnetic field in a next generation nEDM experiment
at the Paul Scherrer Institut.
An array of scalar Cs OPMs has been used successfully to
determine directional and gradient magnetic field information
via fits to multi-sensor readings \cite{PLB2014}.

Various methods exist for extracting information about the magnetic
field vector components, including the spin exchange relaxation free
(SERF) magnetometers \cite{Xia2006AMMEG1} operating at $|\ve{B}_0|=0$
and whose intrinsically sensitivity is to one vector component only.
For operation in the offset fields used for neutron magnetic
resonance, we focus on conventional OPMs that measure the magnetic
field modulus by detecting the Larmor precession frequency,
$\omega_\mathrm{L} = \gamma |\ve{B}_0|$, where $\gamma$ is the
gyromagnetic ratio of the probed atomic state.
Information about vector components is gleaned by monitoring the
OPM's response to an externally applied oscillating magnetic field
using phase sensitive detection.
In a recently published all-optical variant of that method
\cite{Budker2014AllOptical}, circularly polarized laser beams induce
an effect equivalent to the perturbation field via the vector
light shift \cite{Mathur68}.
Without external modulations, vector information can be inferred using
multiple detection channels~\cite{Lenci2014VectorAMTransient} when
the first and second harmonics of the Larmor precession of an atomic
alignment \cite{weis2006dram} are detected with linearly polarized
light.

Our approach uses multiple circularly polarized laser beams to gain
vector information and thus extends methods pioneered by Fairweather
and Usher \cite{Usher1972VecRb}.
The absorption of circularly polarized light depends linearly on the
projection of the atomic spin polarization on the light's $\ve{k}$
vector.
The precessing atomic polarization modulates the transmitted light
power at $\omega_L$ if $\ve{B}_0$ is not parallel to $\ve{k}$.
This can be used to maintain the condition $\ve{B}_0 \parallel \ve{k}$
in a feedback loop either by changing the direction of $\ve{B}_0$
\cite{Usher1972VecRb} or of \ve{k} \cite{Vershovskii2011}.
In contrast to those vector magnetometer implementations, our system
uses off-line data analysis enabling us to infer the magnetic field
information from free spin precession (FSP) signals.
The FSP method is particularly well suited for the application in nEDM
experiments since it allows for very stable field measurements.

\section{Experimental setup}

\begin{figure}
  \centerline{\includegraphics[width=0.7 \columnwidth]{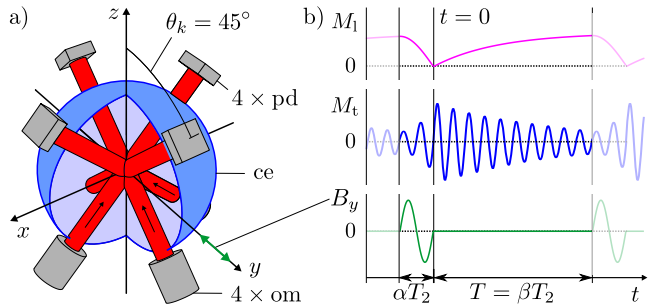}}
  \caption{%
    a) Apparatus schematic: four laser light beams are polarized by the
    optical modules (om), traverse the Cs cell (ce, shown in section) and
    are converted into signals $S_i$ by the photodiodes (pd).
    b) Measurement cycle time structure, repeated every $40\,\mathrm{ms}$:
    The longitudinal magnetization $M_\mathrm{l}$ is created by optical
    pumping.
    Following a $\pi /2$ pulse, $B_y$, of duration $\alpha T_2$, the
    resultant transverse magnetization $M_\mathrm{t}$ gives rise to
    FSP signals (shown here projected onto the $x$ axis).
    The parameters $\alpha, \beta$ measure the length of the $\pi /2$
    pulse and the FSP signal, respectively, in units of $T_2$.
    In each signal analysis, the $t=0$ time origin is reset to the FSP
    start.
  }
  \label{fig:ExperimentalSetup}
\end{figure}

The experiment was performed inside the magnetic shield of the nEDM
experiment at PSI \cite{Baker2011nEDMSearch}, in which a stable
\SI{1}{\micro T} magnetic field was generated by a $\cos \theta$ coil.
The static magnetic field is parametrized as $\ve{B}_0 = B_0 (\sin
\theta \cos \phi, \sin \theta \sin \phi,$ $\cos \theta)$, and was
approximately aligned along the $z$ axis, i.e., $\theta \approx 0$.
Our magnetometer was created to measure the field modulus, $B_0$, and
its direction, i.e., the polar angle, $\theta$, and azimuthal angle
$\phi$.
The magnetometer design is shown in Fig.~\ref{fig:ExperimentalSetup}(a).
Light is generated by an extended cavity diode-laser coupled to a
polarization-maintaining single-mode fiber splitter having three
outputs.
One output feeds a saturated absorption spectroscopy unit, used for
active laser frequency stabilization to the $F{=}4\rightarrow 3$ cesium
$D_{1}$ transition (\SI{894}{nm}).
A~second single-mode fiber guides light to the magnetometer head,
where it is split into four beams which are coupled into short
multi-mode fibers.
At the sensor head, the light from each multi-mode fiber is collimated
and circularly polarized by a linear polarizer and a quarter-wave
plate, mounted in a compact optical module (om).
The power of those beams can be adjusted by rotating an additional
linear polarizer in the om.
The beams traverse an evacuated \SI{45}{mm} diameter glass-cell (ce)
containing a saturated vapor of cesium atoms.
The cell is paraffin coated \cite{Castagna2009Coating} to reduce spin
depolarization during atom-wall collisions.
A~combination of photodiodes (pd) and transimpedance amplifiers
converts the transmitted light power of each beam to a signal $S_i$,
which is digitized with a high resolution sampling system.
The combined noise of the photodiode, preamp, and sampling system is
well below the shot-noise level for the typical light power of
\SI{1}{\micro W} per laser beam.

The magnetometer is operated in pulsed mode, and information
is extracted from the FSP signals.
Figure~\ref{fig:ExperimentalSetup}(b) shows the experimental cycle which
repeats every 40~ms.
The FSP is described using the magnetization $\ve{M}$ associated with
the ensemble average of the atomic spin.
The combined optical pumping by the four laser beams in combination
with $\ve{B}_0$ creates a magnetization $\ve{M}_\mathrm{l}$
longitudinal to $\ve{B}_0$.
A~short magnetic $\pi/2$ pulse along the $y$ direction turns
$\ve{M}_\mathrm{l}$ to a direction transverse to $\ve{B}_0$, creating
$\ve{M}_\mathrm{t}$.
The pulse uses a single sinusoidal period in order to minimize
deadtime ($\alpha$ in Fig.~\ref{fig:ExperimentalSetup}(b)).
The magnetization component perpendicular to $\ve{B}_0$ precesses at
the Larmor frequency, $\omega_\mathrm{L} = \gamma B_0$, where $\gamma
= \SI{3.4986211(4)}{ kHz/\micro T}$ is the gyromagnetic ratio of the
$F{=}4$ cesium ground state.
The light absorption by the cesium atoms depends linearly on the
projection of $\ve{M}$ on the light's $k$-vector~\cite{Dehmelt1957}.
Consequently, the transmitted laser power, measured by a photodiode,
is modulated at $\omega_\mathrm{L}$.
The transverse magnetization component decay (see $M_\mathrm{t}$ in
Fig.~\ref{fig:ExperimentalSetup}(b)), with its effective decay time
$T_2$, is observed as a decreasing modulation amplitude of the
recorded FSP photodiode signal.
During the FSP, the longitudinal magnetization, $M_\mathrm{l}$, is
recreated by optical pumping, such that the next $\pi /2$ pulse can
start the next cycle.
Both the data acquisition system recording the FSP signals and the
function generator producing the $\pi/2$ pulses are synchronized to an
atomic clock.

Using the classical Bloch equation, the recorded signal for each laser
beam can be modeled as
\begin{equation}
  S_i(t) = c_{i} + e^{-\frac{t}{T_2}}
  \left(b_{i} + A_{\mathrm{c},i}\cos{\omega t}
              + A_{\mathrm{s},i}\sin{\omega t} \right)\,.
\label{eq:SignalLinearModel}
\end{equation}
%
%
Both frequency $\omega$ and effective decay time $T_2$ are common
parameters for all simultaneously recorded FSP signals.
The offsets, $c_{i}$ and $b_{i}$, as well as the in-phase,
$A_{\mathrm{c},i}$, and quadrature, $A_{\mathrm{s},i}$, components of
the modulation amplitudes are different for each signal $S_i$.
The $c_{i}$ parameters represent the DC signal offsets and are
proportional to the average light power of beam~$i$.
If the $k$-vector of beam $i$ has a longitudinal component,
the exponential build-up of $M_\mathrm{l}$ contributes to the
absorption it probes.
Assuming that the longitudinal and transverse relaxation rates are
equal allows this contribution to be parametrized by the offsets
$b_i$.
The modulation amplitudes are used to determine the magnetic field
direction.
The $\ve{B}_0$ field magnitude is determined using the estimation of
frequency $\omega$, interpreted as the Larmor frequency.
The magnitude and the extracted field direction are used to
reconstruct the vector magnetic field.

\section{Data analysis}
\label{sec:Data-analysis}

The parameters of Eq.~(\ref{eq:SignalLinearModel}) are extracted with
a precision limited by the Cram\'{e}r-Rao Lower Bound (CRLB)
\cite{rife1974single,Heil2010HeliumMagnetometer}.
The lower limit of the frequency spectral density, $\rho_f$,
calculated with the CRLB for signals with no DC components
($c_{i}=b_{i}=0$) and sampled at a sufficiently high rate ($\gg
\omega_L/2\pi$) is
\begin{equation}
  \rho_{f} \geq \frac{2\,\rho}{\pi \, A \, T_2}
  \sqrt{\frac{\left(\alpha+\beta\right)e^{\beta}\sinh\beta}
             {\cosh2\beta-2\beta^{2}-1}}\,.
\label{eq:VarianceDeltaToZero}
\end{equation}
The length of the FSP signal, $T$, is parametrized in a dimensionless
way by $\beta = T/T_2$ (Fig.~\ref{fig:ExperimentalSetup}(b)), whereas
$\alpha$ measures the dead-time of the $\pi/2$ pulse.
%
%
The spectral density $\rho$ of the photodiode signals is ultimately
limited by shot noise.
The amplitude $A$ is proportional to $M_z$, in the instance before it
is flipped, it thus scales like $A = A_0 \left( 1-e^{-\beta} \right)$.
Given this, Eq.~(\ref{eq:VarianceDeltaToZero}) has a
minimum at $\beta \approx 2.6$.
For technical reasons, the pulse repetition time of $T=40\,\mathrm{ms}$
was chosen to be slightly shorter than the optimum given that
$T_2=20.4(2)~\mathrm{ms}$.
Estimates indicate a $\sim 10\%$ performance gain at $T \approx 2.6~T_2$.

\begin{figure}
\centerline{\includegraphics[width=0.7 \columnwidth]{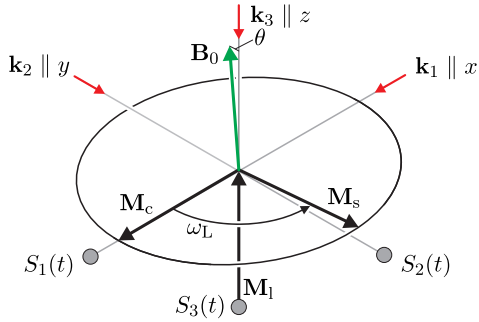}}
\caption{
  The Larmor precession is described by $\ve{M}_\mathrm{c}$ and
  $\ve{M}_\mathrm{s}$, c.f.~Eq.~(\ref{eq:SinCos}), which together
  define a plane perpendicular to $\ve{B}_0$ (here depicted as a
  circle.
The longitudinal magnetization $\ve{M}_\mathrm{l}$ changes only slowly
due to relaxation and optical pumping.
Signal $S_i (t)$ represents the transmitted power of the laser
beam $\ve{k}_i$.  }
\label{fig:Precession}
\end{figure}

Reconstructing the $\ve{B}_0$ vector components is made by
monitoring the $\ve{M}$ component precessing at $\omega_L$. 
By definition, the precession happens in a plane perpendicular to
$\ve{B}_0$, thus the cross product of two vectors in that plane yields
a vector parallel to $\ve{B}_0$.
The method's statistical sensitivity is maximized when the phase
difference of the two vectors is $\pi /2$.
This is achieved by parametrizing the precessing part of $\ve{M}$
by its in-phase and quadrature components as
\begin{equation}
  \ve{M}_\mathrm{t} (t) = \ve{M}_\mathrm{c} \cos \omega_\mathrm{L} t
                        + \ve{M}_\mathrm{s} \sin \omega_\mathrm{L} t\,.
\label{eq:SinCos}
\end{equation}
If the two vectors are known, the $\ve{B}_0$ direction follows as
\begin{equation}
\ve{B}_0 \propto \ve{M}_\mathrm{c} \times \ve{M}_\mathrm{s}.
\label{eq:CrossProduct}
\end{equation}
Measuring $\ve{M}_\mathrm{c}$ and $\ve{M}_\mathrm{s}$ is
straightforward in a three-beam magnetometer with orientations along
the Cartesian coordinates axes (Fig.~\ref{fig:Precession}).
%
%
In such a configuration, the $\cos$ and $\sin$ modulation components
seen by each beam (Eq.~(\ref{eq:SignalLinearModel})) correspond
directly to $\ve{M}_\mathrm{c}$ and $\ve{M}_\mathrm{s}$.

The experimental vector magnetometer reported herein uses four laser
beams (Fig.~\ref{fig:ExperimentalSetup}(a)).
In this configuration, each beam has a \ve{k} component along $z$,
thus contributing to optical pumping provided $\ve{B}_0$ is
approximately oriented along $z$, which is our case.
Each beam probes the projection, $M_i = \hat{k}_i \cdot \ve{M}$,
of the magnetization $\ve{M}$ onto its $k$-vector.
The 3D vector $\ve{M}$ is reconstructed from the four projections,
using the projection matrix $P$
\begin{equation}
\left(\begin{matrix}M_{x}\\
M_{y}\\
M_{z}
\end{matrix}\right) = P\cdot\left(\begin{matrix}M_{1}\\
M_{2}\\
M_{3}\\
M_{4}
\end{matrix}\right) \text{\!, \,}
P=\frac{1}{\sqrt{2}}\left(\begin{matrix}-1 & 1 & 0 & 0\\
0 & 0 & -1 & 1\\
\frac{1}{2} & \frac{1}{2} & \frac{1}{2} & \frac{1}{2} \end{matrix}\right)\,.
\label{eq:PMatrix}
\end{equation}
This reconstruction is advantageous since $M_x$ and $M_y$ are derived
by subtracting two projections, reducing common mode noise.
Since $M_x$ and $M_y$ are the important components for determining the
Larmor frequency, the chosen beam configuration facilitates its high
resolution extraction.
All parts of the signal $S_i (t)$ (Eq.~(\ref{eq:SignalLinearModel}))
that depend linearly on $\ve{M}$ are transformed from the four
projections into a 3D representation using the matrix $P$.
In the low light-power limit, the in-phase and quadrature modulation
amplitudes of $S_i (t)$ are proportional to the DC signal, $c_i$,
detected by photodiode $i$.
Amplitudes $A_{\mathrm{c},i}$ and $A_{\mathrm{s},i}$ are normalized
using $c_i$, to compensate for slight differences in light power and
possible differences in photodiode preamplification factors.
Finally, $P$ converts the normalized amplitudes extracted from signals
$S_i (t)$ to 3D vectors that determine the direction of $\ve{B}_0$
using Eq.~(\ref{eq:CrossProduct}).
Using numerical simulations \cite{sam_phd}, we verified that the resulting
angles $\theta$ and $\phi$ are determined with maximum statistical efficiency.
The magnetic field direction and magnitude can be extracted from the data of only three laser beams.
Given the geometry of the beams in the experiment this is, however, not possible at maximum statistical efficiency.
Correlations in the signals $S_i (t)$ due to the over determined measurement with four laser beams
can be used to verify the normalization factors of the amplitudes \cite{sam_phd}.

The projection matrix $P$ depends on the actual orientation of the laser beams.
Deviations from the assumed orientations lead to systematic errors in the extracted magnetic field orientation $\theta$ and $\phi$.
Those errors depend in a complex way on the orientation of the magnetic field and the direction in which the beam is tilted.
If one laser beam is tilted by an angle $\Delta \alpha$ in a direction that causes the largest errors, it contributes an error of $\Delta \theta = 1/4\, \Delta \alpha$ to the extracted magnetic field orientation.
Tilting all four laser beams in this way is equivalent to tilting the whole sensor by $\Delta \alpha$ which naturally causes an estimation error of $\Delta \theta = \Delta \alpha$.
Tilting all beams in random directions causes a combined error of $\Delta \theta = 1/2\, \Delta \alpha$.
The mechanical construction of the experiment can currently not guarantee a alignment better than $\Delta \alpha=0.004$ rad.

Two estimation methods to extract the parameters of
Eq.~(\ref{eq:SignalLinearModel}) from the digitized signals were
studied: Least-squares fitting, and demodulation.
The least-squares method fits the Eq.~(\ref{eq:SignalLinearModel})
model to the experimental data gained from all beams simultaneously.

The demodulation method uses two-phase lock-in detection with $\cos$
and $\sin$ reference signals at a frequency $\omega_\mathrm{r}$ close
to $\omega_\mathrm{L}$.
This mixes the $\omega_\mathrm{L}$ modulation down to a frequency
close to DC, while noise and other modulations are suppressed by the
low-pass filter \cite{Gaspar2004157}.
The in-phase and quadrature lock-in signals are converted to phase
$\varphi (t_j) \equiv \varphi_j$ and amplitude $A (t_j) \equiv A_j$
for each FSP signal.
The initial modulation amplitude, $A (t=0)$, is extracted by a
least-squares fit of $A (0) \exp{-t/T_2} $ to the time series $A_j$ of
one FSP\@.
The model $\varphi (t) = \varphi(0) + \omega t$ is fitted (with
weighting factors $1/A^2_j$) to the phase signal after correcting for
discrete $2\pi$ steps.
The frequency difference between $\omega_\mathrm{r}$ and
$\omega_\mathrm{L}$ is found using the slope $\omega$.

The in-phase and quadrature modulation amplitudes are
found via $A_{\mathrm{c},i} = A\cos{\varphi(0)}$, and
$A_{\mathrm{s},i} = -A\sin{\varphi(0)}$.
For both methods, the least-squares fits are made simultaneously for
the four FSP signals using one common frequency, $\omega$, and decay
time, $T_2$.

\section{Results}

\begin{figure}
\centerline{\includegraphics*[width=0.7 \columnwidth]{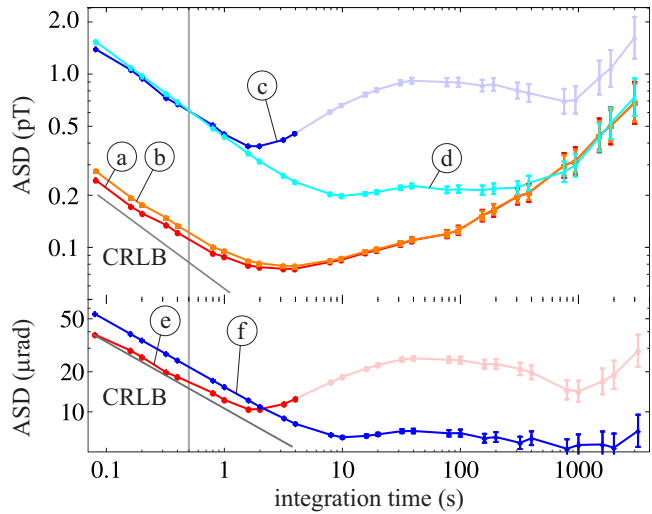}}
\caption{
ASD of the magnetic field magnitude (a, b), the field's z-component
(c, d), and the $B$ direction $\theta$ (e, f) measured with the vector
cesium magnetometer.
The results obtained by least-squares fitting (curves a, c, and e)
generally show better ASD values at short integration times, while the
demodulation method (curves b, d, f) provides better results over
longer integration times.
Curves c and e are affected by systematic errors for integration times
larger than 4 s.
The error bars were calculated according to \cite{Lesage1973}.  }
\label{fig:VCsMStability2014}
\end{figure}

The nEDM experiment requires magnetic field measurements that are
stable over hundreds of seconds at the sub-pT level.
The cesium vector magnetometer statistical errors, as described by the
CRLB, are by far sufficient to reach that goal.
However, systematic errors arising from drifting parameters limit
the long-term stability that this magnetometer can achieve.
To characterize the long-term stability, we measured during 11 hours
under best-case conditions of field stability.
%
Figure~\ref{fig:VCsMStability2014} shows the Allan standard deviation
(ASD) \cite{groeger2006sound} of vector and scalar field measurements
as a function of integration time $\tau$.
Using the estimated FSP parameters and a noise density extracted from
the measured data's Fourier spectrum,
Eq.~(\ref{eq:VarianceDeltaToZero}) yields a CRLB of \SI{81}{fT/
  $\sqrt{\mathrm{Hz}}$} for the field magnitude.
For $\tau < \SI{1}{s}$, the ASD plots show the expected improvement
proportional to $\tau^{-1/2}$.
The least-squares fitting possesses a higher statistical efficiency,
which is visible as a $12\%$ smaller ASD\@.
This difference disappears for $\tau > \SI{1}{s}$ where the ASD is no
longer limited by statistical processes.
For the longer integration times, the ASD is limited by magnetic field
drifts and magnetometer instabilities, thus, the ASD represents the
limit for the magnetometer stability.
The magnitude can be measured with an uncertainty smaller than
$\SI{300}{fT}$ for integration times ranging from $\SI{80}{ms}$ to
$\SI{1000}{s}$.
A~best sensitivity of $\SI{75.2}{fT}$ is achieved at $\tau =
\SI{4}{s}$, which corresponds to a relative sensitivity of
$7.6\times10^{-8}$.

Curves c and d in Fig.~\ref{fig:VCsMStability2014} show the ASD for
the field's $z$ component (longitudinal) measurement.
This shows that the values extracted using the demodulation
method are more stable than those from least-squares fitting for
long integration times.
This happens because the least-squares fitting does not model the
second harmonic of the Larmor modulation, $2\omega_\mathrm{L}$, while
the demodulation method is insensitive to it due to the low-pass
filter.
The integration times for which the $z$ component can be measured with
an uncertainty smaller than $\SI{300}{fT}$ range up to $\SI{1000}{s}$,
but start at $\SI{2}{s}$, due to the larger statistical errors.
This increase in statistical uncertainty is due to using amplitudes
which cannot be estimated as precisely as the Larmor frequency.

Figure~\ref{fig:VCsMStability2014} e~and~f show the ASD of $\theta$,
characterizing the direction $\ve{B}_0$, as derived from the estimated
vector components.
The ASD of $\theta$, estimated using least-squares fitting, scales
statistically for $\tau < \SI{1}{s}$ and complies very well with the
CRLB calculated using error propagation from the estimated amplitudes.
Using demodulation, the resolution of $\theta$ reaches \SI{6.4}{\micro
  rad} for $\tau = \SI{10}{s}$, and does not change significantly
until $\tau = \SI{2000}{s}$.
The ASD of $\phi$ behaves similarly, but with larger uncertainties
since the measurement was made near the degenerate case of
$\theta {=} 0$.

\section{Conclusion and discussion}

The presented magnetometer achieves high sensitivity both in magnitude
and field direction.
Upper limits on processes that limit the stability of the magnetometer
readings are derived from the ASD plots and show high sensitivity
is maintained even at integration times of \SI{1000}{s}.
This value is probably limited by drifts of the $B_0$ field components
in the present nEDM experiment.
Further studies will try to distinguish between instabilities
intrinsic to the magnetometer and external field drifts by using
several magnetometer modules.

In contrast to other recently published vector magnetometers
\cite{Budker2014AllOptical} the presented approach does not degrade
the scalar resolution when extracting vector information.
Consequently, it achieves an order of magnitude better scalar
resolution while being able to resolve the $B_0$ direction with
$\delta \theta < \SI{10}{\micro rad}$ for integration times ranging
beyond \SI{2000}{s}.
This makes the presented approach an ideal choice for applications
that use long integration times.
For our future nEDM apparatus it is planned to use an array of vector
Cs magnetometers in order to monitor the $B_0$ field and its
gradients.
Scaling to multiple sensors is aided by the low needs on
laser power and the efficient data processing possible in the
demodulation mode.

The magnetometer presented here requires calibration in order to
provide absolute field direction information.
However, the accuracy of its absolute field magnitude information may
be limited---as discussed by Grujic et al in~\cite{Grujic2015a}---at the several
10~pT level since $\vec{k}\not\perp\vec{B}_0$.
The demonstrated stability at long integration times is a necessary
step for the future development of such calibration procedures.
With the stability proven, the detailed studies of device construction
systematics (e.g., perturbations to the values in $P$
(Eq.~(\ref{eq:PMatrix})) and device alignment to an external
coordinate system will permit calibration of the device, thus moving
it from being a field stability measurement system to a true field
measurement system.

A~remaining disadvantage of this approach is the non-`magnetically
silent' $\pi/2$ spin manipulation pulse, which can perturb the
environment under study.
A~straightforward way to overcome this is the use of Bell-Bloom
pumping, currently under development within our collaboration
\cite{Grujic2015a}.
A~combination of these two methods is being pursued to provide a
sensitive and magnetically silent vector magnetometer for our
future nEDM search.

\section*{Acknowledgments}

The authors are grateful for financial support from the Deutsche
Forschungsgemeinschaft in the context of the projects BI 1424/2-1 and
/3-1 as well as from the Swiss National Science Foundation,
projects 144473, 149211, and 157079.
The Polish collaborators acknowledge the National Science Centre,
Poland, for the grant No.~UMO-2012/04/M/ST2/00556 and the support by
the Foundation for Polish Science--MPD program, co-financed by the
European Union within the European Regional Development Fund.
The LPC Caen and the LPSC acknowledge the support of the French Agence
Nationale de la Recherche (ANR) under reference ANR-09BLAN-0046.
E.~W.~acknowledges support as a Ph.D. Fellow of the Research Foundation
Flanders
This work is part of the Ph.D. thesis of S.~A.~\cite{sam_phd}.


\begin{thebibliography}{10}
\newcommand{\enquote}[1]{``#1''}

\bibitem{Bloom1962}
A.~L. {Bloom}, \enquote{{Principles of operation of the rubidium vapor
  magnetometer},} Appl. Opt. \textbf{1}, 61 (1962).

\bibitem{Budker2007}
D.~Budker and M.~Romalis, \enquote{Optical magnetometry,} Nat. Phys.
  \textbf{3}, 227--234 (2007).

\bibitem{Bison2003}
G.~Bison, R.~Wynands, and A.~Weis, \enquote{A laser-pumped magnetometer for the
  mapping of human cardiomagnetic fields,} Appl. Phys. B-Lasers O.
  \textbf{76}, 325--328 (2003).

\bibitem{Wyllie20122247}
R.~Wyllie, M.~Kauer, R.~Wakai, and T.~Walker, \enquote{Optical magnetometer
  array for fetal magnetocardiography,} Opt. Lett. \textbf{37}, 2247--2249
  (2012).

\bibitem{Xia2006AMMEG1}
H.~Xia, A.~Ben-Amar~Baranga, D.~Hoffman, and M.~V. Romalis,
  \enquote{Magnetoencephalography with an atomic magnetometer,} Appl. Phys.
  Lett. \textbf{89}, 211104 (2006).

\bibitem{Sander2012}
T.~H. Sander, J.~Preusser, R.~Mhaskar, J.~Kitching, L.~Trahms, and S.~Knappe,
  \enquote{Magnetoencephalography with a chip-scale atomic magnetometer,}
  Biomed. Opt. Express \textbf{3}, 981--990 (2012).

\bibitem{Altarev1996}
I.~Altarev, Y.~Borisov, N.~Borovikova, A.~Egorov, S.~Ivanov, E.~Kolomensky, M.~Lasakov, V.~Nazarenko, A.~Pirozhkov, A.~Serebrov, Y.~Sobolev, E.~Shulgina, V.~Lobashev,
  \enquote{Search for the neutron electric dipole moment,} Phys. Atom. Nucl.
  \textbf{59}, 1152--1170 (1996).

\bibitem{Knowles2009nEDMMagnetometer}
P.~Knowles, G.~Bison, N.~Castagna, A.~Hofer, A.~Mtchedlishvili, A.~Pazgalev,
  and A.~Weis, \enquote{Laser-driven cs magnetometer arrays for magnetic field
  measurement and control,}  Nucl. Instrum. Meth A \textbf{611}, 306--309 (2009).

\bibitem{Smiciklas2011}
M.~Smiciklas, J.~M. Brown, L.~W. Cheuk, S.~J. Smullin, and M.~V. Romalis,
  \enquote{New test of local lorentz invariance using a
  $^{21}\mathrm{Ne}\mathrm{\text{-}}\mathrm{Rb}\mathrm{\text{-}}\mathbf{K}$
  comagnetometer,} Phys. Rev. Lett. \textbf{107}, 171604 (2011).

\bibitem{Peck2012PRA}
S.~K. {Peck}, D.~K. {Kim}, D.~{Stein}, D.~{Orbaker}, A.~{Foss}, M.~T. {Hummon},
  and L.~R. {Hunter}, \enquote{{Limits on local Lorentz invariance in mercury
  and cesium},} Phys. Rev. A \textbf{86}, 012109 (2012).

\bibitem{PLB2014}
S.~Afach, C.~Baker, G.~Ban, G.~Bison, K.~Bodek, M.~Burghoff, Z.~Chowdhuri,
  M.~Daum, M.~Fertl, B.~Franke, P.~Geltenbort, K.~Green, M.~van~der Grinten,
  Z.~Grujic, P.~Harris, W.~Heil, V.~H\'{e}laine, R.~Henneck, M.~Horras,
  P.~Iaydjiev, S.~Ivanov, M.~Kasprzak, Y.~Kerma{}dic, K.~Kirch, A.~Knecht,
  H.-C. Koch, J.~Krempel, M.~Kuzniak, B.~Lauss, T.~Lefort, Y.~Lemi\`ere,
  A.~Mtchedlishvili, O.~Naviliat-Cuncic, J.~Pendlebury, M.~Perkowski,
  E.~Pierre, F.~Piegsa, G.~Pignol, P.~Prashanth, G.~Qu\'em\'ener, D.~Rebreyend,
  D.~Ries, S.~Roccia, P.~Schmidt-Wellenburg, A.~Schnabel, N.~Severijns,
  D.~Shiers, K.~Smith, J.~Voigt, A.~Weis, G.~Wyszynski, J.~Zejma, J.~Zenner,
  and G.~Zsigmond, \enquote{A measurement of the neutron to 199{H}g magnetic
  moment ratio,} Phys. Lett. B \textbf{739}, 128--132 (2014).

\bibitem{Budker2014AllOptical}
B.~Patton, E.~Zhivun, D.~C. Hovde, and D.~Budker, \enquote{All-optical vector
  atomic magnetometer,} Phys. Rev. Lett. \textbf{113}, 013001 (2014).

\bibitem{Mathur68}
B.~S. Mathur, H.~Tang, and W.~Happer, \enquote{Light shifts in the alkali
  atoms,} Phys. Rev. \textbf{171}, 11--19 (1968).

\bibitem{Lenci2014VectorAMTransient}
L.~Lenci, A.~Auyuanet, S.~Barreiro, P.~Valente, A.~Lezama, and H.~Failache,
  \enquote{Vectorial atomic magnetometer based on coherent transients of laser
  absorption in rb vapor,} Phys. Rev. A \textbf{89}, 043836 (2014).

\bibitem{weis2006dram}
A.~Weis, G.~Bison, and A.~S. Pazgalev, \enquote{Theory of double resonance
  magnetometers based on atomic alignment,} Phys. Rev. A \textbf{74}, 033401
  (2006).

\bibitem{Usher1972VecRb}
A.~J. Fairweather and M.~J. Usher, \enquote{A vector rubidium magnetometer,}
  J. Phys. E. Sci. Instrum. \textbf{5}, 986 (1972).

\bibitem{Vershovskii2011}
A.~Vershovskii, \enquote{Project of laser-pumped quantum mx magnetometer,}
  Tech. Phys. Lett. \textbf{37}, 140--143 (2011).

\bibitem{Baker2011nEDMSearch}
C.~Baker, G.~Ban, K.~Bodek, M.~Burghoff, Z.~Chowdhuri, M.~Daum, M.~Fertl,
  B.~Franke, P.~Geltenbort, K.~Green, M.~van~der Grinten, E.~Gutsmiedl,
  P.~Harris, R.~Henneck, P.~Iaydjiev, S.~Ivanov, N.~Khomutov, M.~Kasprzak,
  K.~Kirch, S.~Kistryn, S.~Knappe-Gruneberg, A.~Knecht, P.~Knowles, A.~Kozela,
  B.~Lauss, T.~Lefort, Y.~Lemiere, O.~Naviliat-Cuncic, J.~Pendlebury,
  E.~Pierre, F.~Piegsa, G.~Pignol, G.~Quemener, S.~Roccia,
  P.~Schmidt-Wellenburg, D.~Shiers, K.~Smith, A.~Schnabel, L.~Trahms, A.~Weis,
  J.~Zejma, J.~Zenner, and G.~Zsigmond, \enquote{The search for the neutron
  electric dipole moment at the paul scherrer institute,} Phys. Proc.
  \textbf{17}, 159--167 (2011).

\bibitem{Castagna2009Coating}
N.~Castagna, G.~Bison, G.~Di~Domenico, A.~Hofer, P.~Knowles, C.~Macchione,
  H.~Saudan, and A.~Weis, \enquote{A large sample study of spin relaxation and
  magnetometric sensitivity of paraffin-coated cs vapor cells,} Appl. Phys. B \textbf{96}, 763--772 (2009).

\bibitem{Dehmelt1957}
H.~G. Dehmelt, \enquote{Modulation of a light beam by precessing absorbing
  atoms,} Phys. Rev. \textbf{105}, 1924--1925 (1957).

\bibitem{rife1974single}
D.~Rife and R.~Boorstyn, \enquote{Single tone parameter estimation from
  discrete-time observations,} IEEE T. Inform. Theory  \textbf{20}, 591--598 (1974).

\bibitem{Heil2010HeliumMagnetometer}
C.~Gemmel, W.~Heil, S.~Karpuk, K.~Lenz, C.~Ludwig, Y.~Sobolev, K.~Tullney,
  M.~Burghoff, W.~Kilian, S.~Knappe-Grüneberg, W.~Müller, A.~Schnabel,
  F.~Seifert, L.~Trahms, and S.~Bae{\ss}ler, \enquote{Ultra-sensitive magnetometry
  based on free precession of nuclear spins,} Eur. Phys. J. D \textbf{57}, 303--320 (2010).

\bibitem{sam_phd}
S.~Afach, \enquote{Development of a cesium vector magnetometer for the neutron
  {EDM} experiment,} Ph.D. thesis, ETH-Z\"{u}rich (2014).

\bibitem{Gaspar2004157}
J.~Gaspar, S.~F. Chen, A.~Gordillo, M.~Hepp, P.~Ferreyra, and C.~MarquÃ©s,
  \enquote{Digital lock in amplifier: study, design and development with a
  digital signal processor,} Microprocess. Microsy. \textbf{28}, 157
  -- 162 (2004).

\bibitem{Lesage1973}
P.~Lesage and C.~Audoin, \enquote{Characterization of frequency stability:
  Uncertainty due to the finite number of measurements} IEEE T. Instrum. Meas. \textbf{22}, 157--161
  (1973).

\bibitem{groeger2006sound}
S.~Groeger, G.~Bison, P.~Knowles, and A.~Weis, \enquote{A sound card based
  multi-channel frequency measurement system,} Eur. Phys. J.-Appl. Phys. \textbf{33}, 221--224 (2006).

\bibitem{Grujic2015a}
Z.~Gruji\'{c}, P.~Koss, G.~Bison, and A.Weis, \enquote{A sensitive and accurate
  atomic magnetometer based on free spin precession,} Eur. Phys. J. D
  \textbf{69}, 135 (2015).

\end{thebibliography}
\end{document}